# Open-source Pulseq sequences on Philips MRI scanners

Thomas H. M. Roos[1], Edwin Versteeg[1], Dennis W. J. Klomp[1], Jeroen C. W. Siero[1], Jannie P. Wijnen[1]

[1]Department of Radiology, University Medical Center Utrecht, Utrecht, The Netherlands

**Correspondence**

Thomas Roos, University Medical Center Utrecht, Department of Radiology, Room Q.02.4.52 (HP E.01.132), Heidelberglaan 100 (PO Box 85500), 3584CX Utrecht, the Netherlands.

Email: t.h.m.roos-2@umcutrecht.nl

ORCID: https://orcid.org/0000-0002-3465-7794

## Abstract

**Purpose:** This work aims to address the limitations faced by researchers in developing and sharing new MRI sequences by implementing an interpreter for the open-source MRI pulse sequence format, Pulseq, on a Philips MRI scanner.

**Methods:** The implementation involved modifying a few source code files to create a Pulseq interpreter for the Philips MRI system. Validation experiments were conducted using simulations and phantom scans performed on a 7T Achieva MRI system. The observed sequence and waveforms were compared to the intended ones, and the gradient waveforms produced by the scanner were verified using a field camera. Image reconstruction was performed using the raw k-space samples acquired from both the native vendor environment and the Pulseq interpreter.

**Results:** The reconstructed images obtained through the Pulseq implementation were found to be comparable to those obtained through the native implementation. The performance of the Pulseq interpreter was assessed by profiling the CPU utilization of the MRI spectrometer, showing minimal resource utilization for certain sequences.

**Conclusion:** The successful implementation of the Pulseq interpreter on the Philips MRI scanner demonstrates the feasibility of utilizing Pulseq sequences on Philips MRI scanners. This provides an open-source platform for MRI sequence development, facilitating collaboration among researchers and accelerating scientific progress in the field of MRI.

**Introduction**

Traditionally, MRI sequences are developed and implemented in MRI vendor-specific environments. This makes it challenging for researchers to write and compile new sequences while MRI vendors may be reluctant to share all of their MR scanner source code. Furthermore, it complicates the interchangeability of MRI sequences between researchers using different software versions, especially across various platforms from different MRI vendors. The latter makes it difficult to reproduce MRI sequences accurately across different MR scanners. Moreover, while specific sequences may benefit research, they may not always become available as a product brought to market by the MRI vendor, which therefore cannot be distributed among other institutes, which could hinder scientific progress.

To address these limitations, several open-source MRI sequence frameworks have been designed that are hardware- and software-independent[1,2,3,4,5]. One of these frameworks is Pulseq. Pulseq allows researchers to implement MRI sequences in MATLAB or Python and import these sequences into other (simulation) software or MRI scanners[1,6]. Pulseq has been supported by three major MRI vendors, Siemens[1], Bruker[1] and GE[2] that allow for either high-level translation of Pulseq into their native MRI sequence format, or function similar enough on the lower-level to make an interpreter possible. Other vendors, like Philips, have used a different architecture where interfacing to the Pulseq is less trivial, which may have caused this platform not to be supported to date. In the development platform that Philips offers, called GOAL, the different waveforms and properties of sequence objects, such as gradients and RF (radio frequency) pulses, must be predefined and assigned to a fixed set of compiled objects. These objects can only be minimally varied during execution, most likely to reduce the performance impact of more extensive changes.

However, in recent years, the digital platforms of Philips MRI scanners have become so fast that we hypothesized that more extensive runtime modifications are in reach. We have investigated the use of such runtime object modifications, to allow for extensive runtime changes to waveforms and properties of sequence objects, as a means to implement a Pulseq interpreter for the Philips MRI system. Using cartesian and spiral readouts as the most versatile MRI sequences typically executed, we demonstrate successful performance of open source Pulseq sequences on a Philips MRI system.

## Methods

**Pulseq implementation**

To achieve the required flexibility within the Philips platform, one option would be to predefine many different GOAL objects and toggle them on and off during the sequence execution, as that is one of the allowed runtime modifications. However, this method has the disadvantage of not creating the necessary flexibility, and more complex Pulseq sequences may not be compatible. To achieve the necessary flexibility for a 100% compatible Pulseq interpreter, only a few GOAL objects need to be used, which will be internally modified during sequence execution.

Within the default research agreement with the MRI vendor (Philips), researchers are able to collaborate using the web-based MR-Paradise environment and exchange code in the provided repository. In our study, we have generated a branch of the software release version R5.4, modifying 6 source code files to set up one RF (radio frequency) object, three GR (gradient) objects, one SY (trigger) object, and one AQ (acquisition) object in a dummy sequence. During repeated executions of the dummy sequence, the objects are modified to recreate the Pulseq sequence block by block. These few files will be made available through the MR-Paradise repository, for all research sites of the vendor, which can be accessed through the MR-Paradise forum post linked through [openmr.nl](openmr.nl).

The modified source code files have been compiled using the Philips pulse program environment and copied to the patch folder of the scanner. The 8 channel transmit and 32 channel receive coil (Nova Medical, USA) is plugged into the system and a 10cm diameter spherical phantom is placed in its center and all together are moved to the isocenter of the 7T Achieva MRI scanner (Philips, Best, The Netherlands). A dummy patient is registered in the MRI system, and a default spectroscopy FID (free induction decay) sequence is selected, with the newly introduced 'Pulseq enable' flag set active. When this sequence is run, the Pulseq interpreter takes full control of the 7T Achieva MRI scanner.

**Validation experiments**

The Pulseq interpreter was validated over two different types of sequences, using simulations and scans performed on a 7T Achieva MRI system. First, the gradient echo example sequence available on the Pulseq github was simulated[7]. Here we compared the observed sequence and waveforms of the Philips graphical pulse sequence viewer, to the intended waveforms plotted by Pulseq directly. Subsequently, the same sequence was executed on the MRI scanner, to qualitatively assess the performance of the sequence and of the image reconstruction. Secondly, we performed the spiral sequence example available on the Pulseq github[8], since it contains arbitrary gradient waveforms, and verified the gradient waveforms produced by the scanner using a 16-channel field camera[9] (Skope, Switzerland).

Finally, to assess the potential performance impact, we profiled the CPU utilization of the MRI spectrometer using TimeDoctor II (Nexperia, The Netherlands) when running the aforementioned sequences in the native vendor environment, as well as the Pulseq interpreter.

**Data reconstruction**

The raw k-space samples of the gradient echo scans, acquired using both the native vendor environment and the Pulseq interpreter, were exported from the scanner and read into MATLAB using Reconframe (Gyrotools, Switzerland). The raw data was sorted using either the vendor provided acquisition labels for the native scan, or using the Pulseq sequence file for the Pulseq scan, and then transformed into the image domain by performing a 2D inverse Fourier transformation. The final reconstructions were obtained by combining all receive coils with the RSS (root sum squares) method.

## Results

The 6 adapted source code files were compiled without warnings and loaded onto the MRI system using the research account. The MRI system started with the only warning that a patched system was observed, and that the MRI vendor cannot be held responsible for unintended use. After confirming and accepting the warning, full access to the MRI console was provided. The 'Pulseq enable' switch could be selected in the MRI console environment and after hitting the start scan button, the Pulseq sequence takes over control of the MRI spectrometer.

**Simulations**

Simulations of the MRI spectrometer running the different Pulseq demo sequences resulted in vendor-specific logs that have been visualized through the usage of the accompanying vendor-provided graphical waveform viewer. Figure 1 depicts the simulated gradient echo pulse sequence, which matched exactly to the requested sequence and waveforms, as defined in the Pulseq sequence file. The spiral sequence, and specifically the arbitrary gradient waveforms it contains, are runtime changes to the internal GOAL objects that are normally not supported by the vendor. Therefore these arbitrary gradient waveforms do not show up correctly in the vendor-provided simulation and visualization tools.

**Field measurements**

To validate the gradient waveforms of the spiral sequence, the sequence was run on the MRI scanner and it was measured with the field camera (Dynamic Field Camera, Skope, Switzerland). The field camera could pick up the intended spiral trajectory as shown in figure 2.

All trajectories were measured, up to the level of the final gradient crusher, which caused spins in the field probes to dephase to the level that SNR was insufficient to assess the field. The requested gradients in the Pulseq sequence file are shown next to the gradients captured by the field camera in figure 3, and show only small discrepancies that are expected with a spiral sequence[10,11].

**Image reconstructions**

Figure 4 depicts reconstructed images of the same gradient echo sequence, obtained through the native implementation and Pulseq implementation. Image contrast and SNR the two reconstructed images are comparable. The inhomogeneities are caused by the inhomogeneous $B_1^+$ field at 7 Tesla and the receive profile of the 32-channel head coil.

**Performance profiling**

When observing the MRI spectrometer CPU utilization percentages, as listed in table 1, the spiral sequence only showed minimal resource utilization. The cartesian gradient echo with short $T_E$ and $T_R$ (2.0ms and 4.2ms respectively) demonstrated higher resource utilization. For both scan types, the CPU load on the MRI spectrometer did significantly increase when performing the Pulseq implementation, but all measured values are below 50%.

## Discussion

We have successfully demonstrated the implementation of an interpreter for the open-source MRI pulse sequence format, Pulseq, onto a Philips MRI scanner. Using publicly available example sequences, MRI images could be obtained with the Pulseq environment on the Philips 7T MRI scanner. The current computational performance of the MRI system is more than sufficient to adapt the gradient and RF objects during run-time by the Pulseq code, as shown by the spiral gradient waveforms which consume 15% of its available CPU time.

While a sequence with short and densely filled Pulseq blocks only consumed 44% of the CPU, it would have cost less CPU load (19%) when using the native control by the Philips console. Other, even more resource-intensive sequences, may hit the CPU limit with the current implementation of our Pulseq interpreter for Philips scanners. This could likely be partially mitigated by optimizing the implementation for performance, but some performance overhead as compared to the native methods is expected.

All Pulseq sequences written to date have used Siemens (dwell) timing for defining the temporal grid of gradients (10μs) and RF pulses (1μs). The presented runtime is capable of adopting these timings, except for those related to arbitrary gradient waveforms. These must be supplied with Philips hardware dwell timing (6.4μs) at this time, but these different timings are supported by the Pulseq format.

The current interpreter for the Philips MRI platform requires more dead time around RF and ADC objects than the interpreters for other vendors. These conditions are defined by the drivers of the MRI scanner itself. Therefore, Pulseq sequences may need slight modifications to create the necessary time intervals required by the Philips MRI scanner.

While Pulseq is a widely used open-source platform, several other open-source platforms exist[1,2,3,4,5]. While our implementation was tuned for Pulseq, an almost identical approach is envisioned for the merging of the other frameworks to the platform.

As Philips provides a substantial part of the MRI scanners in the world, it is worthwhile to make the Pulseq available to this vendor as well. With the 4 largest MRI vendors providing access to the dominant part of the MRI market, multicenter trials using the different MRI are in reach. An next step would be to find a framework that facilitates clinical approval status for using such open-source consoles.

**Acknowledgements**

The authors gratefully acknowledge funding from the Dutch Research Council (NWO), as this publication is part of the project "Silent MRI with the speed of CT and richer metabolic information than PET" (with project number 18361) of the research programme "Talent Programme Vidi TTW 2019" which is (partly) financed by the (NWO).

The authors would also like to thank the Spinoza Centre for Neuroimaging (Amsterdam, The Netherlands) for their provided computational facilities.

Open-source Pulseq sequences on Philips MRI scanners                    11

# Figures

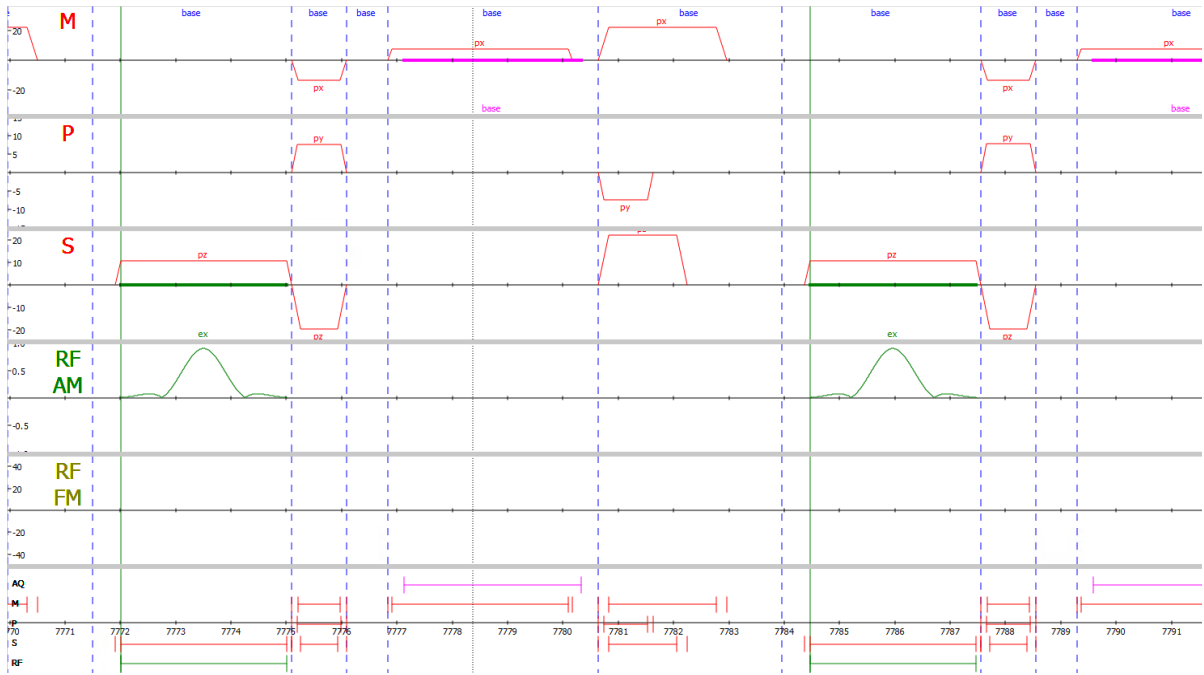

**Figure 1:** Pulse sequence generated from MRI spectrometer simulation, as visualized by the vendor-provided graphical viewer, of the Pulseq interpreter performing the gradient echo example sequence file[7].

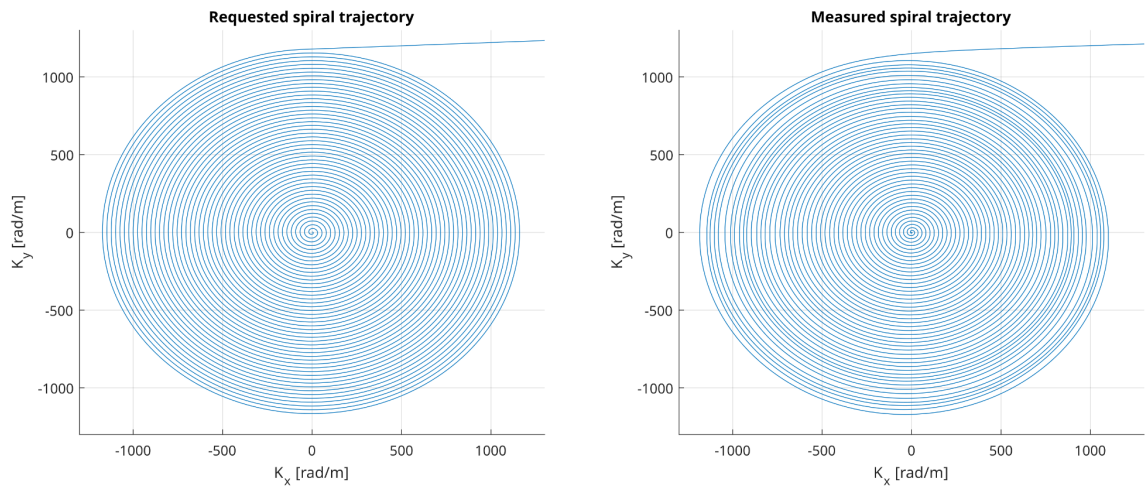

**Figure 2:** k-space encoding trajectory of a spiral readout, both as requested in the Pulseq sequence definition *(left)* and as captured by the field camera *(right)*.

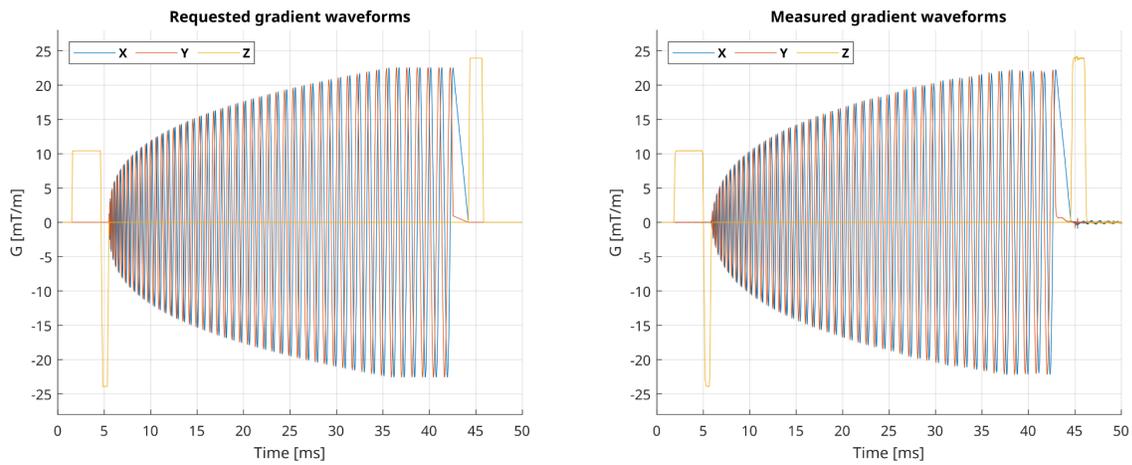

**Figure 3:** Gradient waveforms of a spiral readout, both as requested in the Pulseq sequence definition *(left)* and as measured gradients captured by the field camera *(right)*.

Open-source Pulseq sequences on Philips MRI scanners 15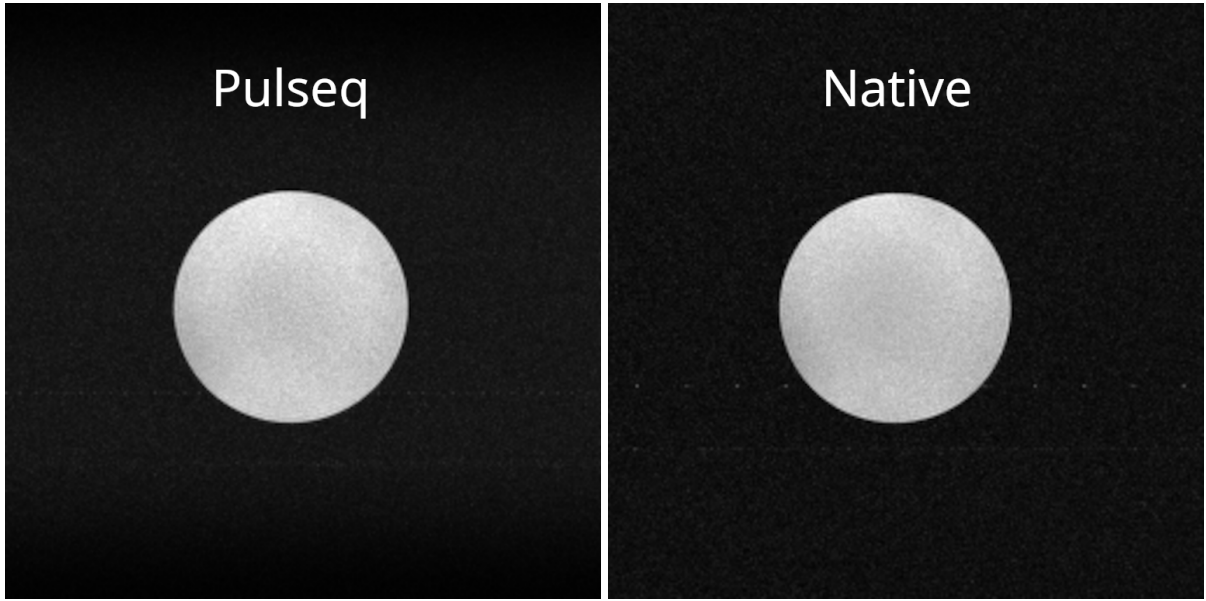

**Figure 4:** Reconstructed images from the Pulseq gradient echo example sequence *(left)* and an native gradient echo (FFE) sequence *(right)* with sequence parameters matched to the Pulseq implementation.

| Scan \ Implementation | Native | Pulseq *(added code)* |
|---|---|---|
| Cartesian (GRE) | 19.15% | 44.24% *(6.42%)* |
| Spiral | 9.03% | 14.63% *(0.11%)* |

**Table 1:** CPU load of the MRI spectrometer during two types of sequences (cartesian and spiral), using either the native implementation *(middle)* or Pulseq implementation *(right)*. Note that the added code takes a relatively small fraction of the workload.